\documentclass[12pt,amsmath,amssymb]{revtex4}
\usepackage{graphicx}
\usepackage{dcolumn}
\usepackage{bm}
\usepackage{wrapfig}
\usepackage{amsmath,amsfonts,amssymb,bm,ulem}
\usepackage[usenames]{color}

\begin{document}

{\center{\large{\bf SUPPORTING ONLINE MATERIAL}}}

\vspace{3mm}

\title{Decoherence in Quantum Magnets: Methods}

\author{ S. Takahashi, I. S. Tupitsyn, J. van Tol, C. C. Beedle,
D. N. Hendrickson and P. C. E. Stamp }


\maketitle

Here we provide details of the methods used in our
high-frequency pulsed ESR experiments, and in the theoretical
analysis of environmental decoherence contributions.

\vspace{3mm}

{\bf (i) EXPERIMENTAL METHODS:} We investigated two slab-shaped
single crystals of Fe$_8$ magnetic molecules (SMMs). These Fe$_8$
crystals were synthesized using the original method developed by
Weighardt et al. \cite{Weighardt84}. ESR measurements
were carried out using a 240 GHz continuous-wave (cw)/pulsed ESR
setup built at the National High Magnetic Field Laboratory (NHMFL),
in Tallahassee FL, USA~\cite{vantol05, morley08}. The system consists of a
40 mW cw solid-state source at 240 GHz, quasioptical ESR bridge,
superheterodyne detection system and 12.5 tesla superconducting
magnet, and employs a $^4$He flow cryostat to vary sample
temperature down to $\sim 1.2$ K. For control of the sample
orientation relative to the external magnetic fields, the sample
crystal was placed in a rotating sample holder consisting of a
teflon rotating rod and nonmagnetic gears. The sample angle can be
changed by rotating a control rod connecting from room temperature
to the gear in the sample holder \cite{vantol05}.

To measure the spin decoherence time $T_2$, we employed a Hahn echo
sequence ($\pi/2 - \tau - \pi - \tau - echo)$ where the delay time
$\tau$ is varied \cite{SpinEcho}. $\pi/2$ and $\pi$ are square
pulses whose width (typically $200$ ns to $300$ ns long) was
optimized for the maximum echo signals. This width corresponds to
approximately $0.5$ Gauss of the excitation bandwidth, much smaller
than the ESR linewidth of the lowest lying Fe$_8$ transition. The
decoherence time $T_2$ was extracted from the decay rate of the echo
area, which was well fit by a single exponential $\exp (-2 \tau /
T_2)$. The temperature dependence of $T_2$ at frequency of $240$ GHz
was then measured between $1.2-2K$, with the external magnetic field
aligned either along the hard (${\bf H} || \hat{x}$) or intermediate
(${\bf H} || \hat{y}$) crystal axes. Above $2$ K, $T_2$ became too
short to give spin echoes within the time resolution of the pulsed
ESR spectrometer.

For the angle-dependent ESR measurement, the sample crystal was
carefully mounted in a rotating sample holder, allowing us to vary
the direction of the external magnetic field along the xy-plane. The
orientation of the crystallographic axes was initially identified
through X-ray diffraction measurements, and then the low-$T$
determination of the actual sample orientation was made using cw
ESR measurements at 20 K. To accurately determine the sample
orientation, we rotated the sample crystal through $360^o$ relative
to the direction of the magnetic fields, and fit the measured
resonance fields to the calculated values as a function of angle
with the Fe$_8$ spin Hamiltonian given in eqtn. (\ref{HS10}) below.
We show here an example of such a fit, for sample $2$ (see Fig. 1).
As a result of these fits, we determined that the direction of the
external magnetic field in the sample 1 was tilted by $<14^o$ from
the $xy$-plane, whereas that in sample 2 was tilted by only $<3^o$.

\begin{figure}[ht]
\vspace{-2.4cm}
\includegraphics[width=9.5cm]{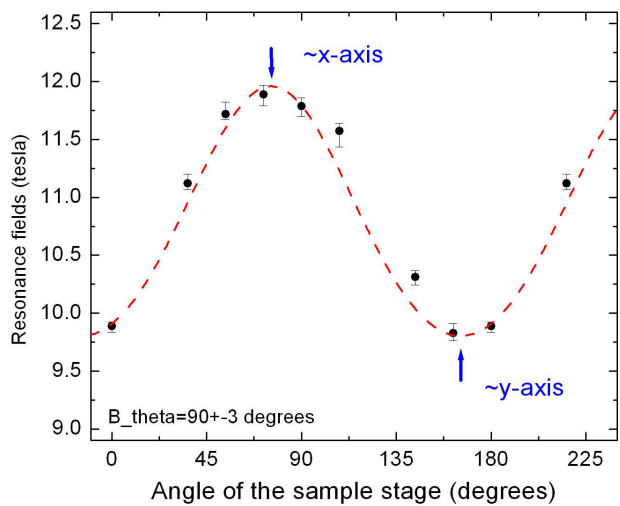}
\vspace{-3.2cm} \caption{
Plot of the resonance fields
measured for Sample 2 as a function of rotation angle of the sample
holder. The dashed red line shows the values calculated from the
Fe$_8$ spin Hamiltonian in eqtn. (\ref{HS10}), described herein. 
}
 \label{Rotn-2}
\end{figure}

\vspace{4mm}

{\bf (ii) THEORETICAL METHODS:} The Fe$_8$ molecule has eight
Fe$^{3+}$ ions ${\bf s}_j$, of spin $5/2$ each, with an electronic
spin Hilbert space of dimension $6^8 \sim 1.7 \times 10^6$. However
below about $10$ K these are locked by superexchange interactions
into a 'giant spin' array, of total spin $|{\bf S}| = 10$. This has
the effective Hamiltonian \cite{BarraCEJ2000}
\begin{equation}
{\cal H}({\bf S}) = -DS^2_z + E S^2_x + K^{\perp}_4
(S^4_{+}+S^4_{-}) - \gamma {\bf S} \cdot {\bf H}_{\perp},
\label{HS10}
\end{equation}
where the easy/hard axes are along $\hat{z}$ and $\hat{x}$
respectively. The Zeeman coupling $\gamma = g_e \mu_B \mu_0$, where
$g_e \approx 2$ and the transverse field $\mathbf{H}_{\perp}$ is
perpendicular to $\hat{z}$.  The parameters of (\ref{HS10}) were
tuned \cite{ST04} to accurately reproduce both the experimental
value of the tunneling gap $2 \Delta_o(\mathbf{H}_{\perp})$ and its
period of oscillations with field \cite{WWSSCI99}, and are $D/k_{\rm
B} = 0.23$ K, $E/k_{\rm B} = 0.094$ K, and $K^{\perp}_4/k_{\rm B} =
-3.28 \times 10^{-5}$ K.

The Hamiltonian (\ref{HS10}) is easily diagonalized numerically. For
all fields used in this experiment, a low-energy doublet clearly
separates off from the higher-energy states - this is the electronic
'qubit'. The energy splitting of this doublet was shown in Fig. 1(b)
of the main text. We write these 2 states as $|\mathcal{S}\rangle$,
$|\mathcal{A}\rangle$, where the lowest energy state is
$|\mathcal{S}\rangle$. Let us define linear combinations
$|\mathcal{Z}_{\pm}\rangle = (|\mathcal{S}\rangle \pm
|\mathcal{A}\rangle)/2^{1/2}$. In the experiment, almost all
molecular spins are in the low-$T$ $|\mathcal{S}\rangle$ state. A
$\pi/2$ microwave pulse then sends these into the
$|\mathcal{Z}_+\rangle$ state, thereby initiating spin oscillations
between $|\mathcal{Z}_{\pm}\rangle$ at a frequency $2\Delta_o/\hbar$
(which may then resonate with the applied ESR signal at frequency
$\Omega_R = 240$ GHz, provided $2\Delta_o/\hbar = \Omega_R$). If we
ignore any field inhomogeneity (which, because of dipolar
interactions, does exist in the sample - see below), these spin
oscillations are equivalent to a uniform spin precession around the
sample along the field directions, {\it i.e.}, to a $k=0$ magnon. A
calculation of environmental decoherence is then equivalent to a
calculation of the coupling of this magnon to the various
environmental modes.

\vspace{5mm}

{\bf Decoherence Rates:} We now calculate the three different
sources of environmental decoherence in the experiment (and their
rather different effect on the ESR linewidth). Note that the results
of these calculations are shown graphically in Fig. 3 of the main
text.

\vspace{3mm}

{\it Nuclear Spins}: In this high field regime, the dimensionless
nuclear decoherence rate reduces to $\gamma^N_{\phi} = E^2_{o}/2
\Delta^2_o$, where $E_o$ is the half-width of the Gaussian multiplet
of nuclear spin states coupled to the qubit states \cite{ST04}. To
evaluate $E_o$ one requires all the hyperfine couplings. These were
previously determined for the Fe$_8$ system, both in low-field
experiments \cite{WW00, Sess01} and theoretically \cite{ST04}. We
extend the theory to high fields by defining the vector
$\mbox{\boldmath $\omega$}_k$, for the $k$-th nuclear spin in the
molecule, with components
\begin{equation}
\omega_k^{\alpha} = {1 \over 2} \sum_j A_{\alpha \beta}^{jk}
(\langle s_j^{\alpha} \rangle^{\cal S} - \langle s_j^{\alpha} \rangle^{\cal A})
 \label{omegak}
\end{equation}
where $\langle s_j^{\alpha} \rangle^{{\cal S},{\cal A}}$ is the expectation value
of the individual spin ${\bf s}_j$, when the molecule is in the
$|{\cal S}\rangle$, $|{\cal A}\rangle$ states. Then $2\mbox{\boldmath $\omega$}_k$
measures the difference
$\mbox{\boldmath $\omega$}_k^{\cal S} - \mbox{\boldmath $\omega$}_k^{\cal A}$
between the hyperfine fields acting on this nuclear spin when the molecule is in the
$|{\cal S}\rangle$, $|{\cal A}\rangle$ states. The nuclear multiplet halfwidth is
then
\begin{equation}
E_o^2 = \sum_k {I_k + 1 \over 3I_k} (\omega_k I_k)^2
 \label{Eo}
\end{equation}
The hyperfine couplings $\{ A_{\alpha \beta}^{jk} \}$ themselves are
essentially independent of field, but the field-dependence of the
qubit states leads to a strong decrease of $E_o$ at high fields.
This is easily found numerically: for these calculations we assumed
natural isotopic concentrations, and found $E_o = 4 \times 10^{-4}$
K for the fields used in this experiment. This gives a quite
negligible contribution to the decoherence rate. It also gives a
very small contribution $E_o$ to the ESR linewidth.

If we do selective isotopic substitution to remove
nuclear spins, we can further decrease the nuclear contribution to
the decoherence rate. Thus substitution of deuterium for protons
decreases $E_o$ from the above value, computed assuming protons at
the 120 different Hydrogen sites, by a factor 3.9, further
decreasing the decoherence rate by a factor 15.2.

\vspace{3mm}

{\it Phonons}: The large tunneling gap means that decoherence from low-energy
acoustic phonons is important. We define $\epsilon_{\alpha \beta} =
(\partial_{\beta} u_{\alpha} + \partial_{\alpha} u_{\beta})/2$ and
$\omega_{\alpha \beta} = (\partial_{\beta} u_{\alpha} -
\partial_{\alpha} u_{\beta})/2$, the symmetric and antisymmetric
parts of the strain tensor, written in terms of the lattice
displacement ${\bf u}({\bf r})$ for a molecule at position ${\bf
r}$. For Fe$_8$ crystals, the mass density $\rho = 1920 kgm^{-3}$
for the natural isotopic concentrations used here, and the unit cell
volume is $V_o = 1956 {\AA}^3$. We assume a phonon spectrum
$\omega_q = q c_s$, with the sound velocity $c_s = (k_B
\Theta_D/\hbar)(V_o/6 \pi^2)^{1/3}$ assumed the same for transverse
and longitudinal phonons; here $\Theta_D = 33$ K is the Debye
temperature \cite{ST04}.

We write the spin-phonon coupling as $\hat{V}_{S,ph} = \sum_i \eta_i
\hat{O}^S_i \hat{O}^{ph}_i$, where the subscript $i$ labels
different irreducible group representations for the lattice and
molecular symmetry groups, and the operators $\hat{O}^S_i,
\hat{O}^{ph}_i$ operate separately on the spin and phonon degrees of
freedom \cite{cal}.  At high fields the dominant terms in
$\hat{V}_{S,ph}$ can be written as
\begin{eqnarray}
\hat{V}^{(X)}_{S,ph} \rightarrow
(\eta_1 \epsilon_{xz} + \eta_2 \omega_{xz})\{ {\hat{S}_x,\hat{S}_z} \}
\nonumber \\
+ (\eta_3 \epsilon_{xy} + \eta_4 \omega_{xy})\{ {\hat{S}_x,\hat{S}_y} \}
\label{V-Sp-X} \\
\hat{V}^{(Y)}_{S,ph} \rightarrow
(\eta_1 \epsilon_{yz} + \eta_2 \omega_{yz})\{ {\hat{S}_y,\hat{S}_z} \}
\nonumber \\
+ (\eta_3 \epsilon_{xy} + \eta_4 \omega_{xy})\{ {\hat{S}_x,\hat{S}_y} \}
\label{V-Sp-Y}
\end{eqnarray}
where the indices $(X),(Y)$ indicate that ${\bf H}$ is parallel to
${\hat x}$ or to ${\hat y}$ respectively and the anticommutators $\{
\hat{A}, \hat{B} \} = \hat{A} \hat{B} + \hat{B} \hat{A}$ are
evaluated at the fields used here. The coupling constants $\eta_j$
can only be determined accurately from experiment. However, simple
estimations \cite{Rotation, Melcher, Dohm} shows that they are $\sim
D/2$, where $D$ is the main easy-axis anisotropy energy.

\begin{figure}[ht]
\vspace{-1.4cm}
\includegraphics[width=9.0cm]{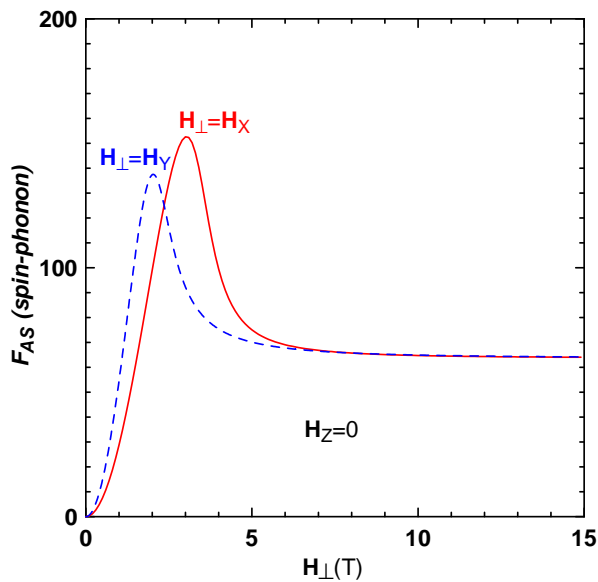}
\vspace{-3.2cm} \caption{
Plots of ${\cal F}_{AS}$
defined in Eqs. (\ref{F-AS2-X}),(\ref{F-AS2-Y}), as a function of
the transverse applied field. The results are shown for
$\mathbf{H}_{\perp}$ along $\hat{x}$ (red) and along $\hat{y}$
(blue).
} 
\label{Sp-Ph-F}
\end{figure}

Since the phonons themselves are only weakly affected by the field,
all field effects come from the field dependence of the qubit states
and $\Delta_o(\mathbf{H}_{\perp})$. The spin-phonon decoherence rate
is then calculated using standard methods \cite{villain,MST06} to
give
\begin{eqnarray}
\tau_{ph}^{-1} \;=\; {\cal F}_{AS} {\Delta_o^3 \over \pi \rho c_s^5
\hbar^4} \coth (\Delta_o/k_B T) \label{tau-ph} \\
{\cal F}_{AS} = \int {d{\Omega} \over \pi} | \sum_i \eta_i
f_i(\Omega) \langle {\cal A} | \hat{O}_i^S | {\cal S} \rangle |^2
\label{F-AS}
\end{eqnarray}
where $d\Omega$ is an infinitesimal solid angle in real space, and
the $f_i (\Omega)$ are rather complicated angular functions coming
from the tensors $\epsilon_{\alpha \beta}$ and $\omega_{\alpha
\beta}$, which are independent of ${\bf q}$. If we now go to the
high field regime, and assume that all the couplings are equal to $D/2$,
we find
\begin{eqnarray}
{\cal F}^{(X)}_{AS} \;\approx\; {D^2 \over 3} (|\langle {\cal A} | \{
{\hat{S}_x,\hat{S}_z} \} |{\cal S} \rangle |^2 + |\langle {\cal A} |
\{ {\hat{S}_x,\hat{S}_y}\} |{\cal S} \rangle |^2) \label{F-AS2-X} \\
{\cal F}^{(Y)}_{AS} \;\approx\; {D^2 \over 3} (|\langle {\cal A} | \{
{\hat{S}_y,\hat{S}_z} \} |{\cal S} \rangle |^2 + |\langle {\cal A} |
\{ {\hat{S}_x,\hat{S}_y}\} |{\cal S} \rangle |^2)
\label{F-AS2-Y}
\end{eqnarray}
These functions are evaluated numerically, by exact diagonalisation;
the results are shown here in Fig. \ref{Sp-Ph-F}.

In general the phonon decoherence rate depends on temperature, but
in our experiments, $k_B T \ll 2\Delta_o$, and so the rate has
already reached its low-$T$ saturation value; this is seen for both
field orientations in Fig. 2 of the main text.

We note immediately that the contribution $\hbar \tau_{ph}^{-1}$ to
the ESR linewidth will be completely negligible.

\vspace{3mm}

{\it Dipolar Interactions}: The dipole field of each molecular spin
leads to a global sample shape-dependent demagnetization field ${\bf
H}_{dm}({\bf r})$, which varies around the sample. Thus the qubit
energy splitting $2\Delta_o ({\bf H}({\bf r}))$ varies around the
sample with the in-plane component of the total local field ${\bf
H}({\bf r}) = {\bf H}_o + {\bf H}_{dm}({\bf r})$. Only in the close
vicinity of 'resonant surfaces' in the sample, where
$2\Delta_o({\bf H}({\bf r})) = \hbar \Omega_R$, can molecules be flipped
by the applied ESR signal. The strength of the ESR signal at a given
applied field ${\bf H}_o$ is then directly proportional to the
number of molecules in this surface, {\it i.e.}, to the density of states
\begin{equation}
{\cal N}({\bf H}_o) \;=\; \int d^3r \; \delta (\hbar \Omega_R -
2\Delta_o({\bf H}_o + {\bf H}_{dm}({\bf r}))
 \label{N-Ho}
\end{equation}
The lineshape will then be complicated, with multiple peaks coming
from the van Hove singularities arising each time a new resonant
surface moves in from the sample boundary, as one varies the
external field. By further varying this field, we sweep the resonant
surfaces through the sample. The edges of the lineshape correspond
to fields where the resonant surfaces are small and near the sample
boundaries, in regions where the demagnetisation field is varying
rapidly with ${\bf r}$. In the middle of the ESR line, the resonant
surfaces occupy a large region near the centre of the sample, where
the demagnetization field is relatively uniform in space; there are
then a large number of molecules in resonance. In this latter case,
the approximation that the tipping pulse corresponds to a uniform
(in space) precession mode, {\it i.e.}, a $k=0$ magnon, is a very good one.
We therefore compare theoretical calculations of the decoherence
with $T_2$ measurements taken at the centre of the line, where the
signal is strongest anyway.

In the present experiment, where $k_B T \ll 2\Delta_o$, only 4-magnon
processes where the $k=0$ mode scatters off a thermally excited
magnon, with two magnon final states, contribute to the decoherence;
decay into three magnons is blocked by kinematic restrictions. It
then follows that the dipole contribution to the dimensionless
decoherence rate $\gamma_{\phi}$ is
\begin{equation}
\gamma_{\phi}^{\rm m} = \frac{2 \pi}{\hbar \Delta_{\rm o}}
\sum_{{\bf q q'}} |\Gamma^{(4)}_{{\bf q} {\bf q'}}|^2 {\cal
F}[\overline{n}_{{\bf q}} ]\; \delta(\omega_0 + \omega_{\bf q} -
\omega_{\bf q'} - \omega_{{\bf q} - {\bf q'}})
 \label{2i2o}
\end{equation}
where ${\cal F}[\overline{n}_{{\bf q}} ]$ is the statistical
distribution function for the magnons, $\Gamma^{(4)}_{{\bf q} {\bf
q'}}$ is the 4-magnon scattering amplitude, and the magnon
dispersion $\omega_{\bf q}$ depends on the applied field and on the
sample geometry \cite{MST06}. We calculated numerically the dipolar
field distributions and the magnon dispersion and decoherence rates
for samples $1$ and $2$, approximating them as parallelipipeds with
the measured aspect ratios. The Fe$_8$ molecules, described by the
full $8$-spins $\mathbf{s}_i = 5/2$ model with the known positions
of each Fe ion \cite{CDC}, were organized in triclinic arrays in the
numerical calculation of the magnon decoherence rates, with angles
and crystallographic axes as in \cite{Weighardt84} and directions of
the anisotropy axes as in \cite{BarraCEJ2000}. We found the width of
the dipolar distributions in both samples to be $\sim 0.1$ T which
is comparable with the width of the experimental ESR peaks. The
magnon contribution to the decoherence rate for this process is
shown in Figs. 2 and 3 of the main text.

\vspace{3mm}

Finally, as noted in the main text, the sample certainly contains
some static disorder, coming from dislocations and defects, as well
as from other lattice strains (which can be induced by a strong
field). The disorder will also cause variation of the Fe$_8$
Hamiltonian parameters between molecules \cite{HillD02, Park02}. The
spin energy levels of each molecule are thus in general shifted, by
different amounts, by this disorder. This will certainly influence
the sample linewidth and lineshape, and thus give an inhomogeneous
broadening contribution to $T_2$. However, provided this disorder
can be considered to be static, it {\it cannot} give any
contribution to the decoherence, which is completely unaffected by
static level shifts. We also note that the experiment is designed so
that any inhomogeneous broadening is canceled by the rephasing
$\pi$-pulse in the Hahn echo sequence. However, any {\it dynamic}
local degrees of freedom, such as 'loose spins' ({\it i.e.}, spin
impurities with energy levels in the energy range below $\sim
10~K$), or mobile charge defects, can cause decoherence. No evidence
is seen in the experiments for such an extrinsic contribution.



\begin{thebibliography}{99}


\bibitem{Weighardt84}
Wieghardt, K. Pohl, K. Jibril, I. \& Huttner, G.
Hydrolysis products of the monomeric amine complex (C$_6$H$_{15}$N$_3$)FeCl$_3$: the structure of the octameric iron(III) cation of
[(C$_6$H$_{15}$N$_3$)$_6$Fe$_8$($\mu_3-$O)$_2$($\mu_2-$OH)$_{12}$]Br$_7$ $\cdot$ 8H$_2$O. {\it Angew. Chem. Int. Ed. Engl.} \textbf{23}, 77 (1984).


\bibitem{vantol05}
van Tol, J. Brunel, L. C. \& Wylde, R. J. A quasioptical transient electron spin resonance spectrometer operating at 120 and 240 GHz.
{\it Rev. Sci. Instrum.} \textbf{76}, 074101 (2005).

\bibitem{morley08}
Morley, G. W. Brunel, L. C. \& van Tol, J. A multifrequency high-field pulsed electron paramagnetic resonance/electron-nuclear double resonance spectrometer.
{\it Rev. Sci. Instrum.} \textbf{79}, 064703 (2008).

\bibitem{SpinEcho}
Hahn, E. L. Spin echoes. {\it Phys. Rev.} \textbf{80}, 580 (1950).

\bibitem{BarraCEJ2000}
Barra, A. L. Gatteschi, D. \& Sessoli, R. High-frequency EPR spectra of [Fe$_8$O$_2$(OH)$_{12}$(tacn)$_6$]Br$_8$:
a critical appraisal of the barrier for the reorientation of the magnetization in single-molecule magnets.
{\it Chem. Eur. J.} \textbf{6}, 1608 (2000).


\bibitem{ST04}
Stamp, P. C. E. \& Tupitsyn, I. S. Coherence window in the dynamics
of quantum nanomagnets. {\it Phys. Rev. B} \textbf{69}, 014401
(2004)

\bibitem{WWSSCI99}
Wernsdorfer, W. \& Sessoli, R. Quantum phase interference and parity effects in magnetic molecular clusters. \textit{Science} {\bf 284}, 133 (1999).


\bibitem{WW00}
Wernsdorfer, W. \textit{et al.} Effects of nuclear spins on the quantum relaxation of the magnetization for the molecular nanomagnet Fe$_8$.
\textit{Phys. Rev. Lett.} {\bf 84}, 2965 (2000);

\bibitem{Sess01}
Sessoli, R. \textit{et al.} Isotopic effect on the quantum tunneling of the magnetization of molecular nanomagnets.
\textit{J. Mag. Mag. Mater.} {\bf 226-230}, 1954 (2001)

\bibitem{cal}
Callen, E. \& Callen, H. Magnetostriction, forced magnetostriction, and anomalous thermal expansion in ferromagnets. \textit{Phys. Rev.} {\bf 139}, A455 (1965).

\bibitem{Rotation}
Melcher, R. L. Experimental verification of first-order rotational effects in the magnetoelastic properties of an antiferromagnet.
\textit{Phys. Rev. Lett.} {\bf 25}, 1201 (1970).

\bibitem{Melcher}
Melcher, R. L. Rotationally invariant theory of spin-phonon interactions in paramagnets.
\textit{Phys. Rev. Lett.} {\bf 28}, 165 (1972).

\bibitem{Dohm}
Dohm, V. \& Fulde, P. Magnetoelastic interaction in rare earth systems.
\textit{Z. Phys. B} {\bf 21}, 369 (1975).

\bibitem{villain}
Hartmann-Boutron,F. Politi, P. \& Villain, V. Tunneling and magnetic relaxation in mesoscopic molecules.
\textit{Int. J. Mod. Phys. B} {\bf 10}, 2577 (1996).

\bibitem{MST06}
Morello, A. Stamp, P. C. E. \& Tupitsyn, I. S. Pairwise decoherence in coupled spin qubit networks. {\it Phys. Rev. Lett.} \textbf{97}, 207206 (2006).

\bibitem{CDC}            Cambridge database: www.ccdc.cam.ac.uk.

\bibitem{HillD02}
Hill, S. {\it et al.} D-strain, g-strain, and dipolar interactions in the Fe$_8$ and Mn$_{12}$ single molecule magnets: an EPR lineshape analysis.
{\it Int. J. of Mod. Phys. B} \textbf{16}, 3326 (2002).

\bibitem{Park02}
Park, K. {\it et al.} Effects of D-strain, g-strain, and dipolar interactions on EPR linewidths of the molecular magnets Fe$_8$ and Mn$_{12}$.
{\it Phys. Rev. B} \textbf{65}, 014426 (2002).





\end{thebibliography}
\end{document}



\title{Decoherence in Quantum Magnets: Theory and Experiment}

\author{ S. Takahashi$^{1,2}$, I.S. Tupitsyn$^{3}$, J. van Tol$^{4}$,
C.C. Beedle$^{5, \star}$,
D. N. Hendrickson$^{5}$, P.C.E. Stamp$^{3}$ }

\affiliation{$^1$ Department of Chemistry, University of Southern
California, Los Angeles, CA 90089, USA \\
$^2$ Department of Physics and Institute for Terahertz Science and
Technology,
University of California, Santa Barbara, CA 93106, USA  \\
$^3$ Pacific Institute of Theoretical Physics, and Department of Physics
and Astronomy, University of British Columbia, Vancouver B.C.,
Canada V6T 1Z1. \\
$^4$ National High Magnetic Field Laboratory, Florida State
University, Tallahassee, FL 32310, USA \\
$^5$ Department of
Chemistry and Biochemistry, University of California, San Diego, La
Jolla, CA 92093, USA }


\begin{abstract}

\end{abstract}

\pacs{03.65.Yz, 75.45.+j, 75.50.Xx}

\maketitle

{\bf Decoherence in Nature has become one of the most pressing
problems in physics. Many applications, including quantum
information processing, depend on understanding it; and fundamental
theories going beyond quantum mechanics have been suggested
\cite{GtHPen, Penrose, Leggett}, where the breakdown of quantum
theory appears as an 'intrinsic decoherence', mimicking
environmental decoherence \cite{PS06}. Such theories cannot be
tested until we have a handle on ordinary environmental decoherence
processes. Here we show that the theory for insulating electronic
spin systems can make accurate predictions for environmental
decoherence in molecular-based quantum magnets \cite{Gattes}.
Experimental understanding of decoherence in molecular magnets has
been limited by short decoherence times, which make coherent spin
manipulation extremely difficult \cite{Taka09,
Bertaina08,Schlegel08,Ardav}. Here we reduce the decoherence by
applying a strong magnetic field. The theory predicts the
contributions to the decoherence from phonons, nuclear spins, and
intermolecular dipolar interactions, for a single crystal of the
Fe$_8$ molecular magnet. In experiments we find that the decoherence
time varies strongly as a function of temperature and magnetic
field. The theoretical predictions are fully verified
experimentally - there are no other visible decoherence sources. Our
investigation suggests that the decoherence time is ultimately
limited by nuclear spins, and can be extended up to about 500
microseconds, by optimizing the temperature, magnetic field, and
nuclear isotopic concentrations. }

Environmental decoherence processes are reasonably well understood
at the atomic scale \cite{atomicDec} (although some poorly
understood noisy sources remain \cite{chuang08}). However both
quantum information processing, and the fundamental tests noted
above, require an understanding of decoherence in larger systems,
where experimental decoherence rates are usually much larger than
theoretical predictions. This discrepancy is usually attributed to
"extrinsic" sources (external noise, uncontrolled
disorder/impurities). We thus need to find systems, with many
degrees of freedom, where extrinsic decoherence can be eliminated,
and where we have a quantitative understanding of other decoherence
sources.

Many insulating electronic spin systems are currently the subject of
intense experimental interest, notably in semiconductor quantum dots
\cite{Hanson07}, nitrogen-vacancy centres in diamond
\cite{Takahashi08,Childress06,Dobro}, and large spin magnetic
molecules \cite{Bertaina08,Schlegel08,Taka09}. In all these systems,
three environmental decoherence mechanisms are involved. The
electronic spins couple locally to (i) phonons (an oscillator bath
\cite{AJL87}); (ii) to large numbers of nuclear spins (a spin bath
\cite{PS00}); and (iii) to each other via dipolar interactions. The
long range of dipolar interactions is a major problem: it makes
quantum error correction more difficult, is theoretically
complicated \cite{MST06}, and is very hard to eliminate
experimentally.


Two great advantages of the Fe$_8$ system \cite{Weighardt84} are
that (i) the interaction strengths are well known, allowing
quantitative predictions; and (ii) it can be prepared with little
disorder and few impurities, reducing the danger of extrinsic
decoherence. The number of relevant environmental degrees of freedom
is very large; depending on isotopic concentrations, there are
$10^{50} - 10^{54}$ nuclear spin levels in each molecule; and the
system couples to a bulk phonon bath. At low temperature, the
electronic spin system behaves as a 2-level quantum bit (qubit),
with a splitting depending strongly on an applied transverse field,
perpendicular to the easy axis $\hat{z}$ (See
Fig.~1(b)). While almost all previous experiments in the
electron spin systems examined the low-field regime, where nuclear
spins make the qubit dynamics incoherent, here we go to high
transverse fields; by suppressing most nuclear spin decoherence, we
unveil the coherent qubit dynamics.

Spin echo measurements, using a Hahn echo sequence \cite{SpinEcho},
were carried out on two single crystals of Fe$_8$ molecules (see
Fig.~1), with natural isotopic concentrations, using a
240 GHz pulsed ESR spectrometer \cite{vantol05, morley08}.
Fig.~1 shows Fe$_8$ ESR signals obtained by
echo-detected field sweep measurements. The resonant peaks are
broadened, with a width $\sim 0.1~T$; the peculiar structure of the
peak when ${\bf H} \parallel \hat{y}$ comes from dipolar
interactions \cite{Supp}.

\begin{figure}[ht]
\includegraphics[width=8cm]{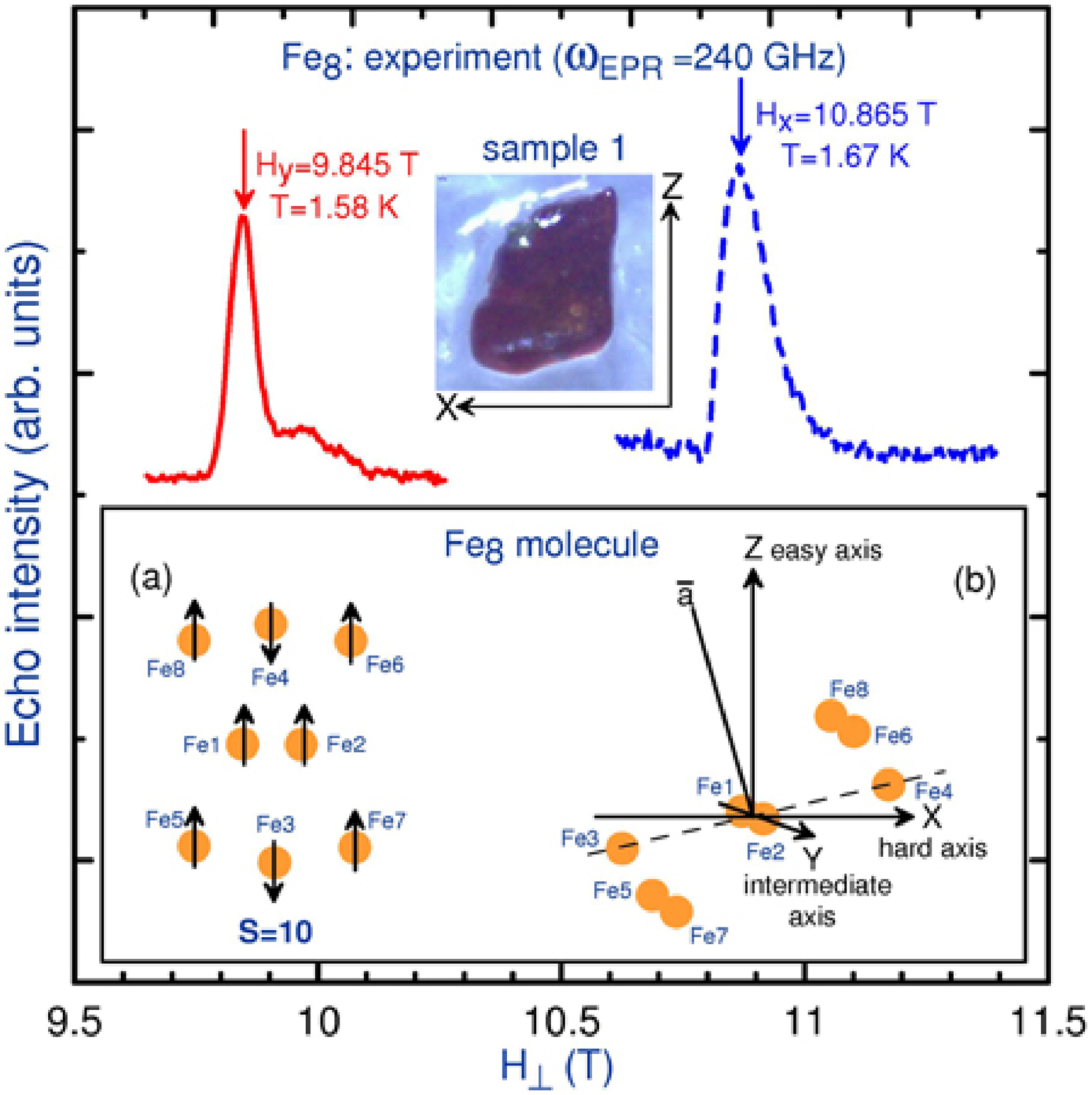}
\includegraphics[width=8cm]{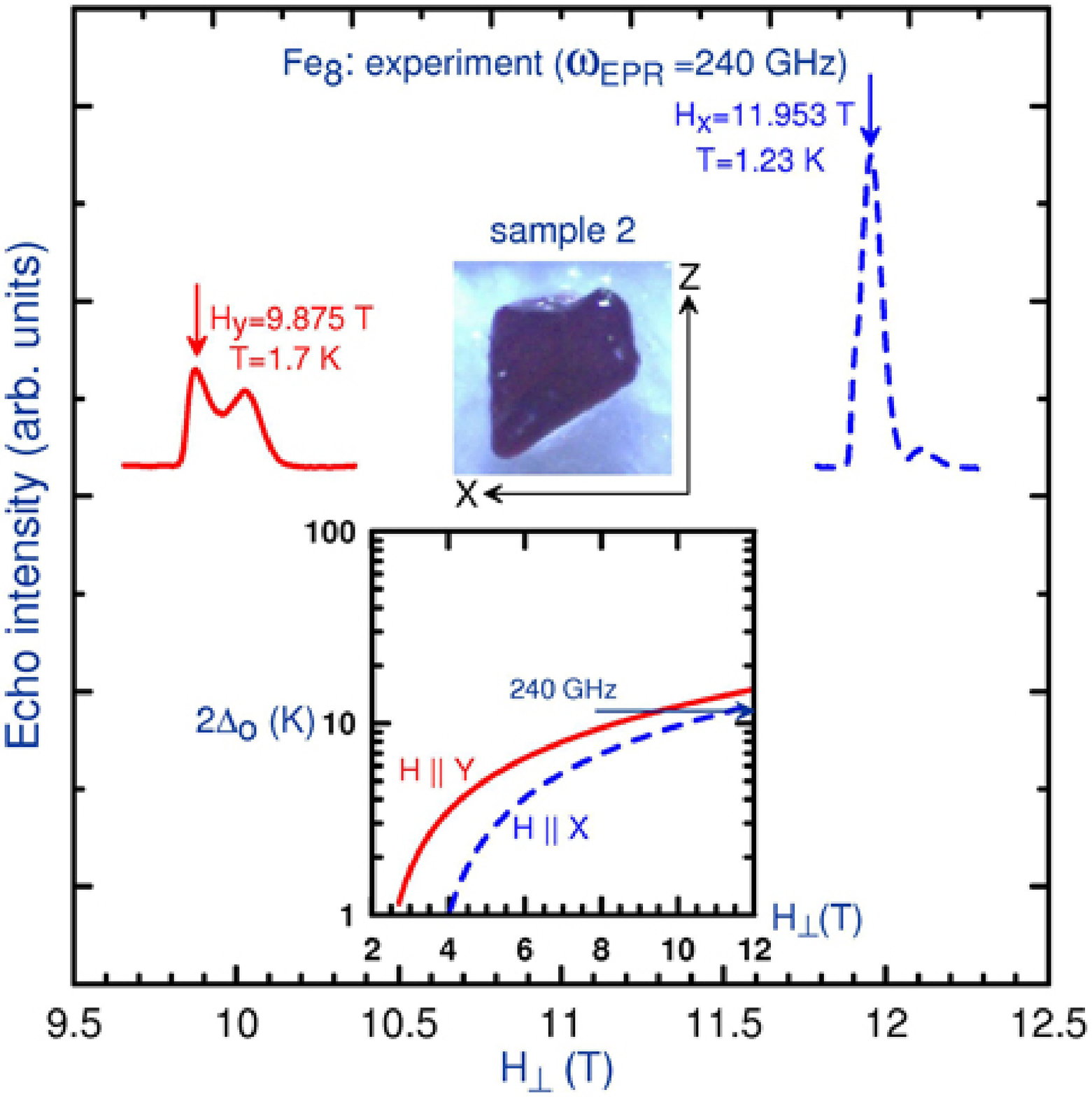}
\caption{
Typical spin-echo detected field-sweep ESR
signals as a function of transverse magnetic field
$\mathbf{H}_{\perp}$ recorded in two different samples, at different
temperatures and orientations of field. (a) For sample 1, Solid red
line - $\mathbf{H}_{\perp} || \hat{y}$, $T = 1.58$ K. Dashed blue
line - $\mathbf{H}_{\perp} || \hat{x}$, $T = 1.67$ K. The sample
dimensions are approximately $z:x:y = 1000 : 700 : 250 \; \mu m^3$.
The lower left inset (i) shows the low-$T$ spin structure of the
Fe$_8$ molecule. The lower right inset (ii) shows the directions of
the easy and hard axes of a Fe$_8$ molecule ($\vec{a}$ approximately
gives the direction of the crystallographic vector $\mathbf{a}$).
(b) For sample 2, Solid red line - $\mathbf{H}_{\perp} || \hat{y}$,
$T = 1.7$ K. Dashed blue line - $\mathbf{H}_{\perp} || \hat{x}$, $T
= 1.23$ K. The sample dimensions are approximately $z:x:y = 900 :
800 : 400 \; \mu m^3$. The lower inset shows the tunneling
splitting, $2 \Delta_o$, as a function of transverse field at
$\mathbf{H}_{\perp} || \hat{y}$ (solid red line) and at
$\mathbf{H}_{\perp} || \hat{x}$ (dashed blue line).
}
 \label{Echo_12n}
\end{figure}

These ESR signals may be understood as follows. The qubit splitting
$2 \Delta_o$ varies with field as shown in the inset to
Fig.~1(b). For fields $H_y = 9.5T, H_x = 0$, or $H_x =
11.3T, H_y = 0$, the electronic Hamiltonian for Fe$_8$ \cite{Supp}
predicts $2 \Delta_o(\mathbf{H}_{\perp}) \approx 11.5 \; K$,
equivalent to our spectrometer frequency of $240$ GHz, implying we
should see resonance peaks at these fields. These predictions are
reasonably well satisfied in both samples. The discrepancies,
discussed in detail in the supplementary material, come from two
sources: (i) sample misorientation (ii) weak high field departures
from the model Hamiltonian \cite{Barra00} used to predict the field
splitting.

In this experiment, the uniform ESR precession mode (a ${\bf k} = 0$
magnon) interacts with its surroundings; to understand the
decoherence time $T_2$, and the ESR lineshapes, we need to look at
the processes contributing to them. For convenience, we define a
dimensionless decoherence rate $\gamma_{\phi} = \hbar/\Delta_o T_2$.
Then the contributions to $\gamma_{\phi}$ are as follows (the full
quantitative discussion, for the 2 samples in this experiment, is in
the supplementary material):

(i) In this high field regime, nuclear spins interact locally with
each molecular spin, and cause decoherence by a 'motional narrowing'
process in which they attempt to entangle with the fast-moving qubit
\cite{MST06,ST04}. The nuclear decoherence rate is $\gamma^N_{\phi}
= E^2_{o}/2 \Delta^2_o$, where $E_o$ is the half-width of the
Gaussian multiplet of nuclear spin states coupled to the qubit; and
the nuclear contribution to the ESR linewidth is just $E_o$. Now in
this experiment, with naturally occurring isotopic concentrations,
$E_o = 4 \times 10^{-4}~K$, whereas $\Delta_o = 5.75~K$. Thus
$\gamma^N_{\phi} \sim 10^{-9}$ is quite negligible, and the nuclear
spin contribution $\sim E_o$ to the linewidth is also very small
compared to the main contributions. Isotopic
substitution of deuterium for the 120 protons in each molecule will
further decrease $\gamma^N_{\phi}$ by a factor 15.2 to
$\gamma^N_{\phi} \sim 7 \times 10^{-11}$.

(ii) The form of the local spin-phonon interaction is determined by
the system symmetry. At high fields this interaction simplifies, and
we find a dimensionless phonon decoherence rate \cite{MST06} given
by $\gamma_{\phi}^{ph} = [( {\cal F}_{AS} \Delta_o^2 ) / (\pi \rho
c_s^5 \hbar^3 )] \coth(\Delta_o / kT)$, where $\rho$ is the sample
density, $c_s$ the sound velocity, and ${\cal F}_{AS}$ is the spin
part of the spin-phonon matrix element \cite{Supp}. While
$\gamma_{\phi}^{ph}$ depends strongly on field, its temperature
dependence is very weak for these experiments. The contribution of
this spin-phonon process to the ESR linewidth is negligible.

(iii) The intermolecular dipole interaction leads to direct coupling
of the ${\bf k} = 0$ ESR precession mode to finite momentum magnons;
these cause strong decoherence. The $k=0$ magnon can in principle
decay spontaneously into multiple magnons, or scatter off thermal
magnons. The dipolar interaction affects the ESR lineshape and the
decoherence rate very differently. The long-range interaction
creates a distribution of demagnetization fields around the sample.
In the case of highly polarized samples, this is strongly
sample-shape dependent, but for annealed samples, it is Gaussian
distributed \cite{vill00,PST04}; in both cases it can be calculated
numerically. The lineshape then reflects the quite broad
distribution of these fields. However, decoherence comes from the
magnon decay process described above, and depends only on the phase
space available at the resonance field for these processes; it can
then be calculated directly from the analytic expression for this
process. At the experimental temperatures, the magnon decoherence
rate $\sim \exp (- 2 \Delta_o / k_B T)$, coming almost entirely from
thermal magnon scattering. This decoherence contribution increases
rapidly with temperature.

(iv) Finally, there can be {\it extrinsic} contributions from
impurities and defects, which typically cause the easy-axis
anisotropy parameter $D$ of the Fe$_8$ Hamiltonian to fluctuate
around the sample (the '$D$-strain' effect \cite{HillD02, Park}).
This will certainly contribute to the ESR linewidth. However these
impurities and defects will {\it not} contribute to the decoherence
rate at all, provided they are static, since then they simply shift
the individual qubit energies. Any impurities or defects with
significant dynamics will cause extrinsic decoherence, which would
then lead to a disagreement between theory and experiment - which is
not seen in the present study.

\begin{figure}[ht]
\includegraphics[width=8cm]{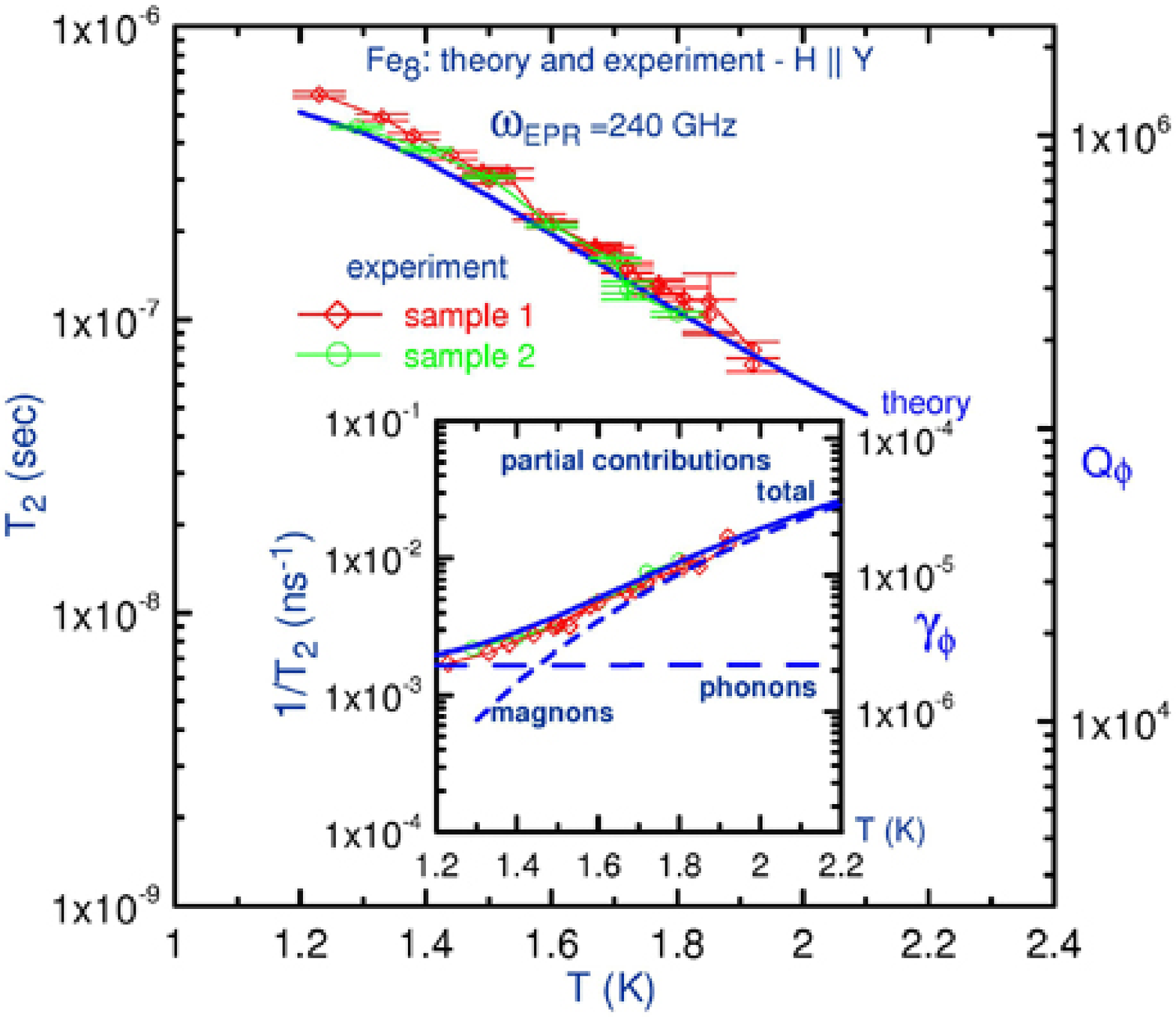}
\includegraphics[width=8cm]{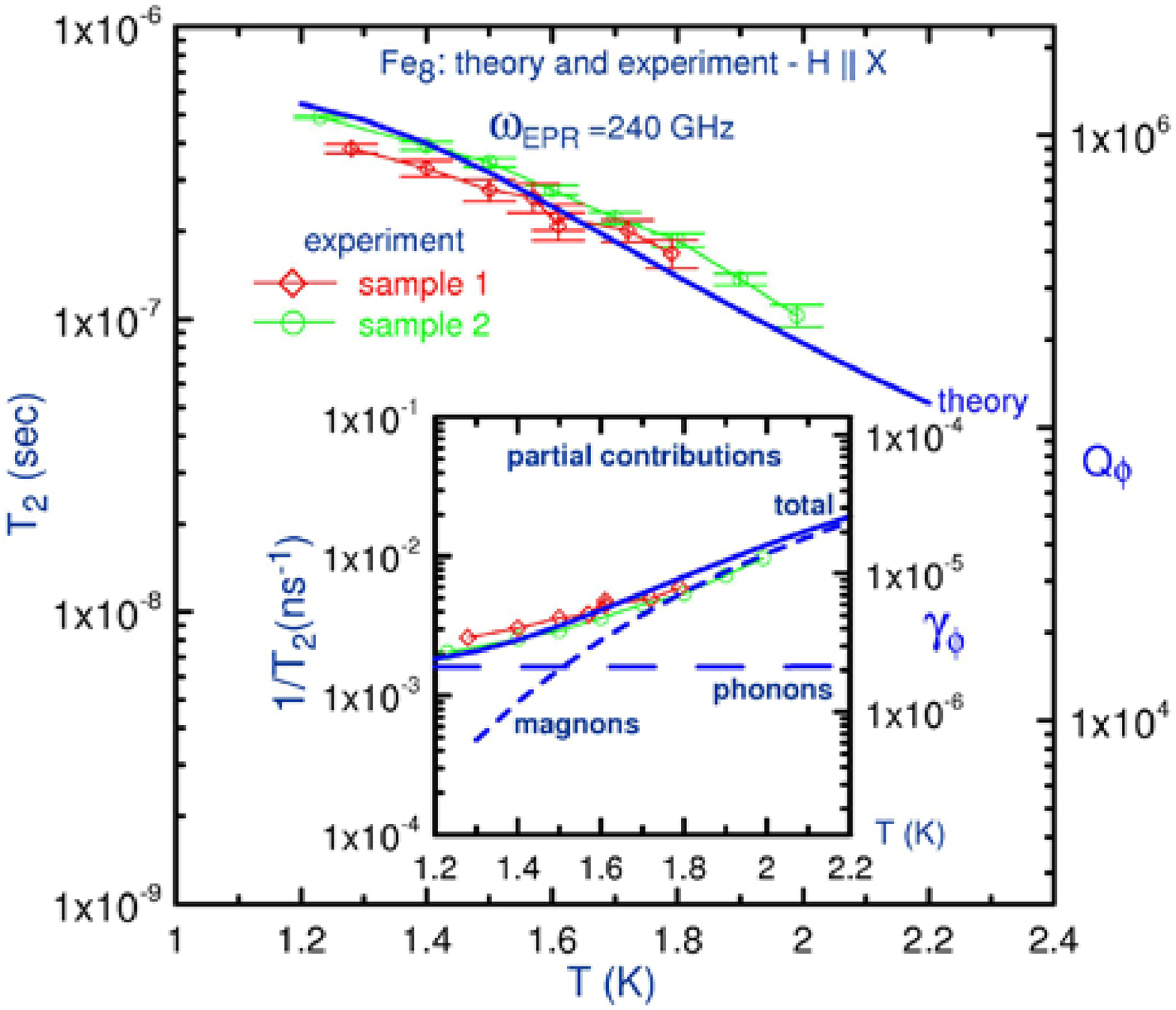}
\caption{
The measured and
calculated decoherence times $T_2$ in samples 1 and 2, as a function
of temperature. (a) Results for $\mathbf{H} || \hat {y}$. Thin line
with diamonds - experiment in sample $1$, $H_y = 9.845$ T. Thin line
with circles - experiment in sample $2$, $H_y = 9.875$ T. Vertical
and horizontal bars represent the experimental errors and
uncertainty in temperature ($\Delta T = \pm 0.05$ K). Thick solid
line - theory including phonon and magnon contributions, $H_y = 9.5$
T. The inset shows the partial contributions calculated for
$T^{-1}_2$ (solid line) from magnons (dashed line) and phonons
(long-dashed line), together with the corresponding experimental
results for the two samples (diamonds and circles). The scale on the
right-hand side of the main figure indicates the decoherence
$Q$-factor $Q_{\phi} = \pi/\gamma_{\phi} = \pi T_2 \Delta_o/\hbar$;
the right-side scale on the inset shows $\gamma_{\phi}$. (b) The
same as for (a), but now for $\mathbf{H} || \hat{x}$. The
experimental curves were measured at $H_x = 10.865$ T (sample 1) and
$H_x = 11.953$ T (sample 2). The theoretical curves are obtained at
$H_x = 11.3$ T. Again, the scales on the right-hand side of the
figure and inset indicate the decoherence $Q$-factor $Q_{\phi}$ and
$\gamma_{\phi}$ respectively, as in (a).
}
 \label{T2_YXn}
\end{figure}

The resulting theoretical decoherence times for Fe$_8$, for each
sample, are presented in Fig.~2, together with the
results of the measurements. The agreement is very good; we
emphasize that apart from the size of the spin-phonon coupling,
which is only known approximately, there are no adjustable
parameters in these fits.

A number of features should be stressed here. First, notice how
different the effects of these mechanisms is on decoherence and on
the the ESR lineshape. The ESR linewidth and lineshape are
completely dominated by static impurity fields and by the spatially
varying dipolar fields. However the decoherence is completely
dominated, at these high fields, by the phonons and dipolar
interaction effects. At lower fields, the nuclear spin decoherence
would also be important, but its effect on the ESR linewidth would
still be negligible. Note also that whereas the dipolar
contributions to the ESR lineshape depends strongly on sample shape,
this shape can only affect the dipolar decoherence near the edges of
this line. In the middle of the line, where the decoherence is
coming from molecules near the centre of the sample, where the field
is homogeneous, one expects no dependence of the dipolar decoherence
on sample shape. This is also what we found in the experiment.

Second, we emphasize how the experiment tests the phonon and dipolar
contributions to the decoherence separately - they
have very different temperature dependences in the regime covered
here, with phonons dominating below $\sim 1.2~K$, and magnons
dominating at higher temperature. We find agreement between theory and
experiment, with no adjustable parameters, across nearly an order of
magnitude in decoherence rate. Thus all decoherence in the
experiment can be accounted for by environmental sources. This
implies we have no measurable extrinsic decoherence here, either
from disorder or dynamic impurities. Nor do we have evidence for any
other contributions, either from "3rd-party decoherence"
\cite{PS06}, or from any of the "intrinsic decoherence" sources
\cite{GtHPen} discussed in the literature.

\begin{figure}[ht]
\includegraphics[width=8cm]{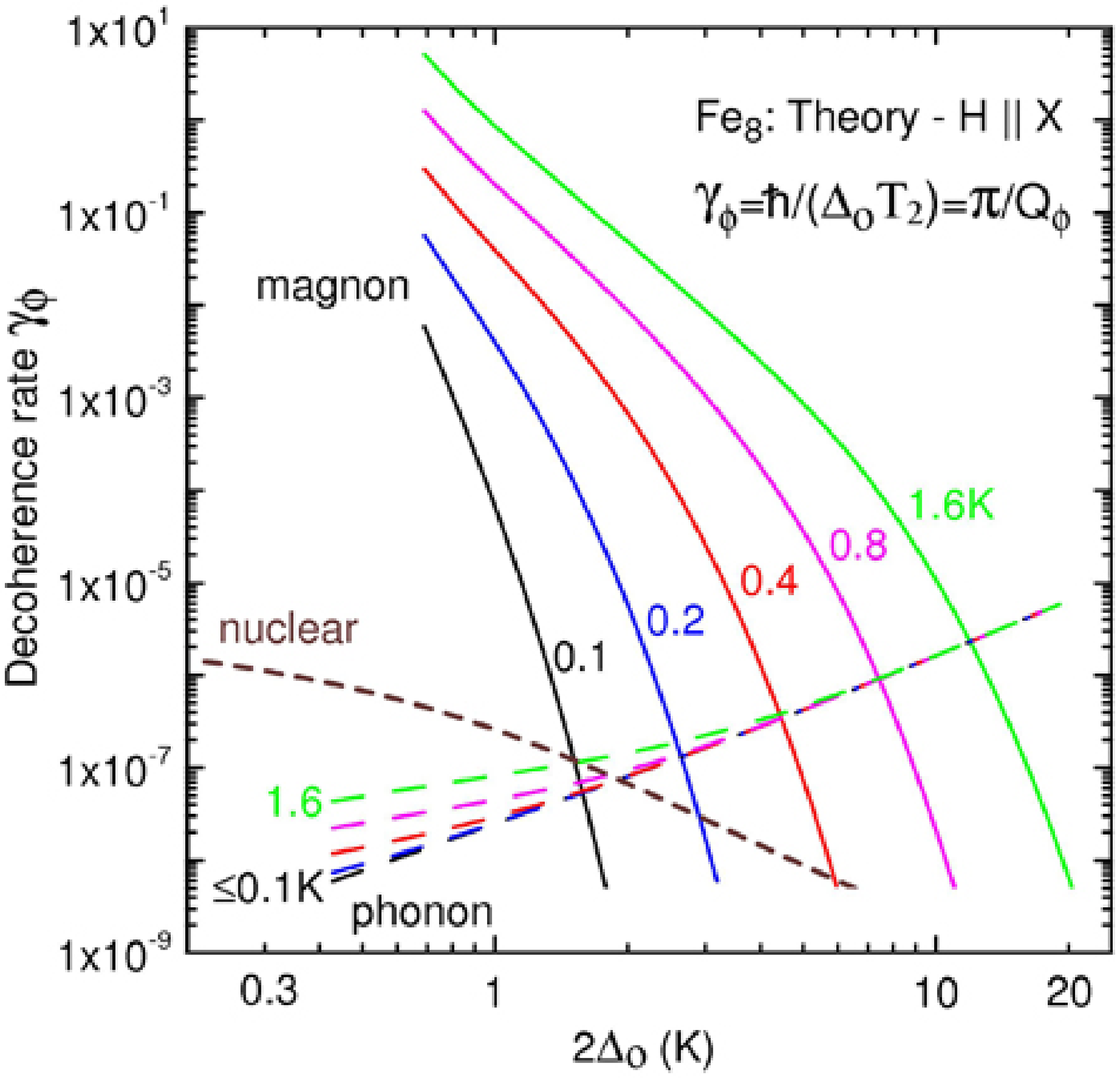}
\includegraphics[width=8cm]{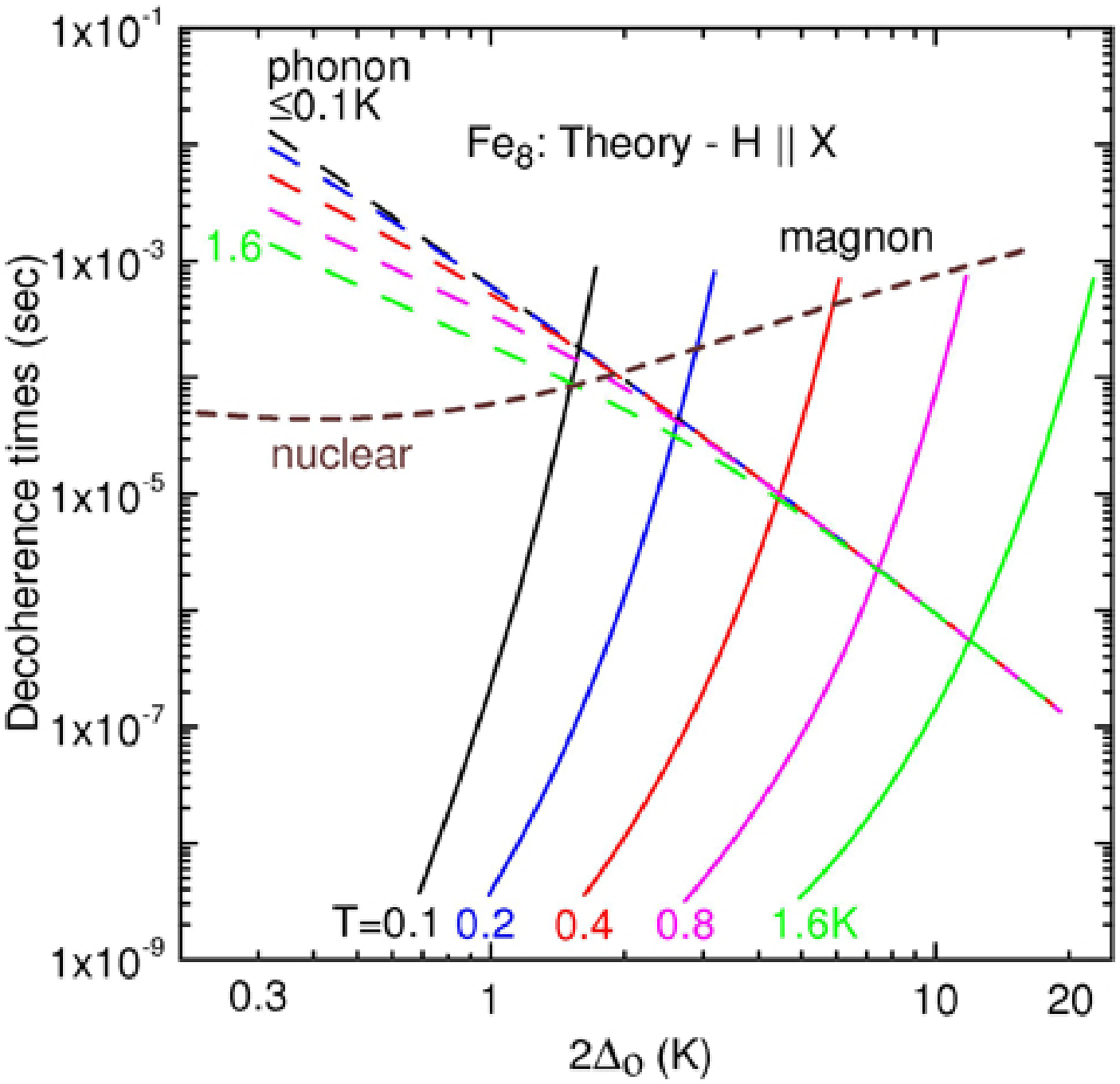}
\caption{
Calculated
contributions to the decoherence coming from the coupling to nuclear
spins, phonons, and magnons respectively. (a) shows the 3 individual
contributions which sum to give the dimensionless decoherence rate
$\gamma_{\phi} = \hbar/T_2 \Delta_o$, as a function of the qubit
splitting in the case ${\bf H}_{\perp} || \hat{x}$. (b) shows the
corresponding three contributions to the decoherence time $T_2$. In
each figure: Brown dashed line - nuclear contribution; Solid lines -
magnon contributions at different temperatures from $0.1 - 1.6~K$;
Long dashed lines - phonon contributions, shown for the same
temperatures as in the magnon case.
}
 \label{deco_rate_time}
\end{figure}

The present investigation suggests that $T_2$ in the Fe$_8$ SMM
system can be increased significantly by optimizing the temperature
and the qubit splitting 2$\Delta_o$. If we extend the calculations
to cover all temperature and field regimes accessible to resonance
experiments, we obtain \cite{Supp} the results shown in Fig.~3. Going to lower temperature allows use of a
smaller ESR frequency 2$\Delta_o$; both these changes strongly
reduce the dipolar and phonon decoherence contributions, giving a
large increase in $T_2$. The optimal decoherence rate comes at the
point where the phonon and nuclear spin decoherence contributions
cross (for $2 \Delta_o \sim 2~K$); provided that $kT < 0.13~K$, so
that the dipolar contribution can be ignored. One then finds that
$\gamma_{\phi} \sim 1.5 \times 10^{-7}$, corresponding to a
decoherence time $T_2 \sim 50 \mu$secs. Any further
increase in $T_2$ would require isotopic substitution, notably of
deuterons in place of the protons; we then find the optimal
decoherence time to be $T_2 \sim 500 \mu$secs, at $2\Delta_o = 0.8K
\equiv 17$ GHz, and $T = 45$ mK. These considerations show the
usefulness of this kind of theory in the optimal design of spin
qubit systems. If quantum mechanics is to be tested on anything but
microscopic scales, it will be essential to continue developing
theory and experiment for systems like the present one, where the
environmental decoherence processes can be understood
quantitatively, and where extrinsic decoherence sources can be
largely eliminated.

\newpage

$\star$ Present address: National High Magnetic Field Laboratory, Florida State
University, Tallahassee, FL 32310 USA.

\newpage

\textbf{Supplementary information}
\\
Supplemental material is linked to the online version of the paper at www.nature.com/nature.

\vspace{3mm}

\textbf{Acknowledgments}
\\
This work was supported by the NSF (DMR-0520481, DMR-0703925), and
the Keck Foundation (S.T. and J.v.T.); and by NSERC, CIFAR, PITP, the
John E. Fetzer Memorial Trust (grant D21-C62), and the Center for
Philosophy and the Natural Sciences, California State University,
Sacramento (I.T. and P.S.). The National High Magnetic Field
Laboratory is supported by the NSF Cooperative Agreement
DMR-0654118, by the State of Florida, and by the DOE.

\vspace{3mm}

\textbf{Author contributions}
\\
S.T. I.T. and P.S. contributed to writing of the manuscript. S.T. I.T. and P.S.
conceived the ESR experiments. The ESR measurement were carried out by S.T. and J.v.T.
The theoretical work has been done by I.T. and P.S.
C.B and D.H. synthesized Fe$_8$ crystals and performed X-ray diffraction measurements.

\vspace{3mm}

\textbf{Author Information}
\\
Reprints and permissions information is available at
www.nature.com/reprints. The authors declare no competing financial interests.
Readers are welcome to comment on the online version of this article at
www.nature.com/nature. Correspondence and requests for materials should be
addressed to S.T. (susumuta@usc.edu) or P.S. (stamp@phas.ubc.ca).